\documentclass[a4paper, oneside, twocolumn, notitlepage, 10pt]{extarticle_ecoc}
\usepackage{ecoc}
\usepackage{comment}

\begin{document}
\selectlanguage{english}    % Standard Language

%-------------------------------------------------- Title -----------------------------------------------------%

%\title{On the Network-wide Benefits of Bidirectional Transmission in Hollow-Core Fiber Networks}%
%\title{Network-Wide Impact of Bidirectional Transmission in Hollow-Core Fiber Networks}
%%\title{Network-Wide Benefits of Selective Bidirectional Transmission in Hybrid Hollow-Core Fiber Networks}
%\title{Network-Wide Benefits of Selective Bidirectional Transmission in Hybrid SMF/HCF Optical Networks}
\title{Selective Deployment of Bidirectional Hollow-Core Fibers in Hybrid SMF/HCF Optical Networks}

%------------------------------------------------- Authors-----------------------------------------------------%

\author{
    Mëmëdhe Ibrahimi\textsuperscript{*}, Giovanni S. Sticca\textsuperscript{*}, Angelo Ferrara, Massimo Tornatore
    %Author\textsuperscript{(1)}, Author\textsuperscript{(1)},
    %Author\textsuperscript{(2)}, Author\textsuperscript{(3)}
}

\maketitle                  % Create title and author

%------------------------------------------ Description of Authors ----------------------------------------------%

\begin{strip}
    \begin{author_descr}

        Politecnico di Milano, corresponding author:
        \textcolor{blue}{\uline{memedhe.ibrahimi@polimi.it}} \\
        \textsuperscript{*}\footnotesize{M. Ibrahimi and G. S. Sticca are co-first authors}

    \end{author_descr}
\end{strip}

% \setstretch{1.1}
%-------------------------------------------------- Footnote -------------------------------------------------------%
\renewcommand\footnotemark{}
\renewcommand\footnoterule{}
%\let\thefootnote\relax\footnotetext{text}

%-------------------------------------------------- Abstract ---------------------------------------------------------%

\begin{strip}
    \begin{ecoc_abstract}
        We investigate selectively deploying bidirectional transmission in hybrid Hollow-Core Fiber (HCF) networks. Upgrading 50\% of links to bidirectional HCF yields at least a 40\% throughput increase compared to unidirectional SMF and captures 85\% of the power consumption reduction of a full unidirectional HCF network upgrade.
        ©2026 The Author(s)
    \end{ecoc_abstract}
\end{strip}

\section{Introduction}
Recent evolution in Hollow-Core Fiber (HCF) technology, especially in Nested Anti-Resonant Nodeless Fiber (NANF), is reaching a significant level of maturity while delivering breakthrough improvements in transmission performance~\cite{Petrovich2025}. Compared to Single Mode Fiber (SMF), %HCF exhibits a significantly lower transmission loss (between 0.08-0.11 dB/km~\cite{Chen_24}, most recently as low as 0.04 dB/km~\cite{YOFC-004})
HCF exhibits lower transmission loss (0.04--0.11 dB/km~\cite{Chen_24, YOFC-004}), ultra-low nonlinearity (up to 3-4 orders of magnitude lower), and 30\% lower transmission latency. 

However, large-scale HCF deployment faces several challenges, primarily the high cost of HCF itself and the fact that low-loss HCF designs typically require a larger cladding diameter compared to SMF~\cite{Poletti:25}. %As HCF deployment is foreseen for green-field scenarios~\cite{Poletti:25} 
Hence, this larger diameter would result in fewer fibers fitting within a standard duct, potentially bottlenecking the total spatial capacity. % of the infrastructure. 
Furthermore, the CapEx required for a complete infrastructure overhaul makes a 100\% network-wide HCF deployment economically prohibitive for most operators in the near-to-medium term~\cite{mem_ondm_25}.

To overcome the duct-size constraint as well as to maximize return on investment, %of the expensive fiber, 
\emph{bidirectional transmission} over a single HCF strand (\emph{BiDi-HCF}) emerges as a compelling architecture. In traditional SMF, same-wavelength bidirectional transmission is severely bottlenecked by Rayleigh backscattering~\cite{10585304}.  
%Moreover, because SMF is remarkably cheap, deploying additional accompanying equipment (e.g., optical circulators and amplifiers) to support BiDi transmission is economically unjustified. 
Conversely, the unique physical properties of HCF, where light travels predominantly in air, result in a Rayleigh backscattering coefficient orders of magnitude lower, i.e., 30 dB lower~\cite{Slavik:22}, than that of SMF. %, i.e., 30 dB lower compared to SMF~\cite{Slavik:22}. 
Given the high CapEx of HCF, the CapEx and OpEx of additional equipment (e.g., optical circulators and amplifiers) are justified by the effective doubling of per-fiber capacity. 
Furthermore, the improved Generalized Signal-to-Noise Ratio (GSNR) in HCF, thanks to low transmission loss and low nonlinearities, enables a paradigm shift in transponder allocation: 
%Furthermore, the improved signal quality in HCF, i.e., high Generalized Signal-to-Noise Ratio (GSNR) thanks to low transmission loss and low nonlinearities, enables a paradigm shift in transceiver allocation: 
BiDi-HCF deployments can %overwhelmingly 
rely on low-cost, energy-efficient 400G/800G ZR+ pluggable modules rather than power-hungry Long-Haul (LH) transponders~\cite{Qiaolun-JOCN}.

To leverage these benefits while managing CapEx, operators can adopt a \emph{selective upgrade} strategy. %Rather than a full network overhaul, creating a hybrid architecture, where only a strategic percentage of links are upgraded to BiDi-HCF while the rest remain legacy unidirectional SMF (Uni-SMF), can unlock substantial performance improvements. 
Rather than a full network overhaul, operators can unlock substantial performance improvements by creating a hybrid architecture, i.e., by upgrading only critical links to BiDi-HCF while retaining the rest as unidirectional SMF (Uni-SMF).
%By selectively deploying BiDi-HCF on critical links, operators can effectively bypass duct capacity bottlenecks and maximize the end-to-end feasibility of ZR+ transponders, maintaining significant capacity and power consumption gains comparable to a 100\% HCF deployment while vastly outperforming the Uni-SMF baseline.

%\textbf{Novelty and State-of-the-Art Positioning:} 
Recent literature has demonstrated the feasibility of BiDi-HCF, showcasing record-breaking capacities, such as 273.6 Tb/s in CLS-band transmission~\cite{Zhang:25}, high-power BiDi transmission over 100 km using 400G ZR~\cite{BiDi-Microsoft-OFC26}, and a system-level analysis in a submarine cable reaching 1 Pb/s~\cite{Poggiolini-BiDi}. 
Meanwhile, network-wide studies have evaluated HCF for low-latency DCI~\cite{OFC24,TNSM-polimi}, high-power amplification~\cite{10926228,11029374}, and network capacity increase with partial HCF deployment~\cite{polimi-OFC25,Joao-OFC2026}. 
However, no prior work has evaluated the network-wide benefits of BiDi-HCF, nor its application in hybrid SMF/HCF networks. Therefore, in this study, we address the following research question: 
\emph{How does the selective deployment of BiDi-HCF influence network-wide throughput, power consumption per Tbps, and transceiver allocation (ZR+ vs. Long-Haul), compared to traditional uniform deployments?}

\section{System Model and Physical Layer Modeling}
To evaluate the network-wide impact of BiDi-HCF under CapEx constraints, we identify three different link-level deployments: % 
\emph{i)} \emph{Uni-SMF}: deployment of a pair of unidirectional SMF; \emph{ii)} \emph{Uni-HCF}: deployment of a pair of unidirectional HCF; and \emph{iii)} \emph{BiDi-HCF}: deployment of a single HCF strand with bidirectional transmission over the same spectrum.

\begin{figure}[t]
    \centering
    \includegraphics[width=1.0\columnwidth]{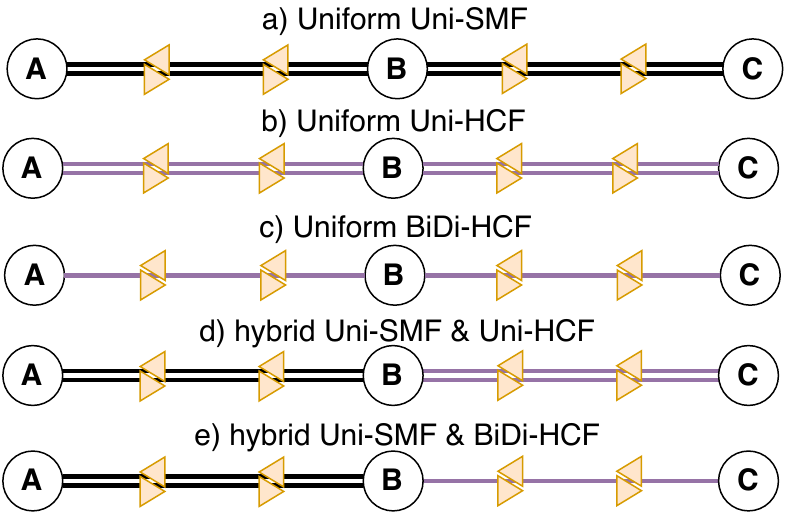}
    \caption{Network-wide deployment scenarios}
    \label{fig:scenarios}
\end{figure}

Based on these link deployments, we define five network scenarios, as illustrated in Fig~\ref{fig:scenarios}: 
\begin{enumerate}
    \item {Uniform Uni-SMF}: All network links (i.e., 100\%) consist of standard Uni-SMF.
    \item {Uniform Uni-HCF}: All network links (i.e., 100\%) consist of standard Uni-HCF.
    \item {Uniform BiDi-HCF}: All network links (i.e., 100\%) consist of a single BiDi-HCF. 
    \item {Hybrid Uni-SMF/Uni-HCF}: A constrained scenario where only a strategic subset of links ($X\%$) is upgraded to Uni-HCF, while the remaining links operate on legacy Uni-SMF. 
    \item {Hybrid Uni-SMF/BiDi-HCF}: Our proposed architecture, where a selective subset of critical links ($X\%$) is upgraded to BiDi-HCF. 
\end{enumerate}
For each network scenario, we solve the Routing, Modulation format, and Spectrum Assignment (RMSA) problem while selecting the optimal transponder type and ensuring lightpath feasibility based on the required GSNR. 

To determine which links are upgraded in the hybrid architectures, we employ a topological ranking strategy based on \emph{edge betweenness centrality}, i.e., we calculate the number of shortest paths between all source-destination node pairs that traverse each link. The network links are then ranked, and the top $X\%$ of links with the highest betweenness centrality are selected for the HCF upgrade (either Uni-HCF or BiDi-HCF). 

Once the physical infrastructure has been selectively upgraded, we evaluate the network's provisioning capabilities. 
Given an initial set of traffic requests (source-destination pairs and data rates), we route and provision lightpaths using a $k$-shortest-path routing heuristic ($k = 3$) with a First-Fit spectrum assignment policy. 
For each candidate path, we evaluate all feasible transmission modes that satisfy the GSNR constraint and fit within the available spectrum. The selection of the transmission mode and of its corresponding transponder type is governed by the following objective function:

\vspace{-10pt}
\begin{footnotesize}
\begin{equation}\label{eq:txp_obj}
%\vspace{-5pt}
    \min \; \mathcal{J} = \alpha \left( \frac{\widetilde{P}_i}{\widetilde{P}_{\max}} \right) + \beta \left( \frac{S_i}{S_{\max}} \right),
\end{equation}
\end{footnotesize}
where $\widetilde{P}_i$ is the power consumption of transponder type $i$, and $S_i$ denotes its spectrum slot occupation. 
To ensure these terms are dimensionless and comparable in magnitude, they are normalized by $\widetilde{P}_{\max}$ and $S_{\max}$, which represent the maximum power consumption and maximum slot occupation across the %portfolio of 
available transponders, respectively.

The weighting factors $\alpha$ and $\beta$ (where $\alpha + \beta = 1$)\footnote{Setting $\alpha=1$ prioritizes energy efficiency, favoring low-power ZR+ pluggables. Conversely, $\beta=1$ prioritizes spectral efficiency, naturally favoring Long-Haul (LH) transponders.} govern the trade-off between reducing power consumption and minimizing the spectrum footprint. % and reducing power consumption.
Table~\ref{tab:txp-table} summarizes the available transponders, detailing their GSNR requirements, spectrum occupation, and power consumption (based on commercial datasheets, e.g.,~\cite{infinera_chm2t_chm1t_ds}).

\begin{table}[t]
    \centering
    \footnotesize
    \caption{Set of available transponders and their physical parameters: Required GSNR ($GSNR_R$) at $0.1~nm$, slot occupation ($S_i$), and power consumption ($\widetilde{P}_i$).}
    \label{tab:txp-table}
    \resizebox{\columnwidth}{!}{%
    \begin{tabular}{|c|c|c|c|}
        \hline \textbf{TXP type} & $GSNR_{0.1nm}$ [dB] & $S_i$ [$n \times 12.5$ GHz] & $\widetilde{P}_i [W]$ \\
        %\hline  TXP type & $GSNR_R$ [dB] & $S_i = n\times12.5$ GHz  & $\widetilde{P}_{\text{TXP}}$    \\
        \hline  LH200 & 12.4 & 6    & 120              \\
        \hline  LH400 & 18.5 & 6    & 164              \\
        \hline  LH600 & 23.7 & 6     & 208             \\
        \hline  LH800 & 24.9 & 8     & 251             \\
        \hline  ZR+400 & 17.2 & 12     & 23             \\
        \hline  ZR+600 & 21.5 & 12     & 30             \\
        \hline  ZR+800 & 23.4 & 12     & 37             \\
        \hline
    \end{tabular}
    }
\vspace{-15pt}
\end{table}

\textbf{Physical layer modeling}. 
We consider an optical network operating in (C+L) band (10~THz total spectrum) 
with a flex-grid spacing of 12.5~GHz. For SMF, we adopt the physical-layer model detailed in~\cite{SemrauSMF}, %which accounts for both Amplified Spontaneous Emission (ASE) noise and non-linear interference (NLI), 
accounting for both ASE and NLI noise, 
assuming an attenuation coefficient of $0.22$~dB/km. For HCF, we utilize the model detailed in~\cite{Poggiolini2022_hcf_transmission}, and an attenuation coefficient of $0.11$~dB/km and an inter-modal interference (IMI) of $-60$~dB/km~\cite{Poggiolini2022_hcf_transmission}. 

For the \emph{BiDi-HCF} scenarios, we evaluate the impact of backscattering power using the BiDi transmission model introduced in~~\cite{10527057} and validated in ~\cite{Zhang:25,Poggiolini-BiDi}. Specifically, the Rayleigh backscattering coefficient in HCF is set to $-100$~dB/m~\cite{Zhang:25}, which is approximately 30~dB lower than that of SMF~\cite{Zhang:25,Slavik:22}. A provisioned lightpath is feasible if its GSNR strictly exceeds the required threshold for the selected transmission mode (i.e., transponder type and modulation format), including a $1$~dB system margin. % to account for operational uncertainties.
\begin{figure}[b]
    \centering
    \vspace{-20pt}
    \includegraphics[width=1.0\columnwidth]{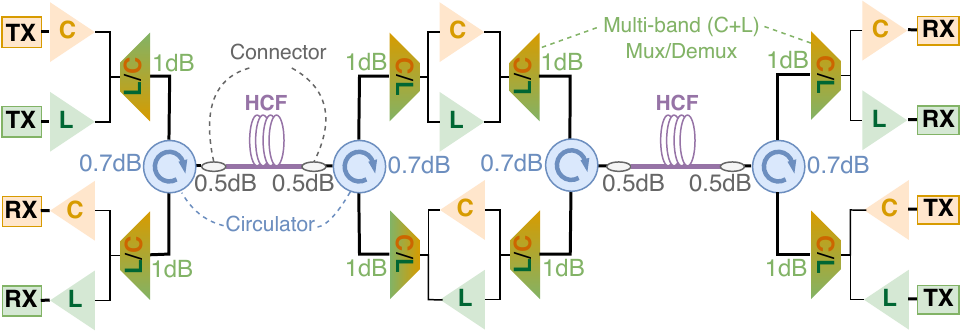}
    \caption{(C+L)-band BiDi-HCF transmission system}
    \label{fig:example-BiDi}
\end{figure}
To visualize the modeled link architecture and its associated lumped losses, Fig.~\ref{fig:example-BiDi} illustrates a two-span link in the \emph{BiDi-HCF} scenario, detailing an optical circulator at both ends of each span to separate the counter-propagating signals and the band MUX/DEMUX to isolate the bands and route them to band-dedicated amplifiers, assuming insertion losses of $0.5$~dB/connector, $0.7$~dB/circulator~\cite{Zhang:25}, and $1.0$~dB/MUX-DEMUX~\cite{Souza-ofc} (for schematic clarity, Tx/Rx losses are omitted).

\begin{figure*}[t]
    \centering
    \includegraphics[width=0.48\textwidth]{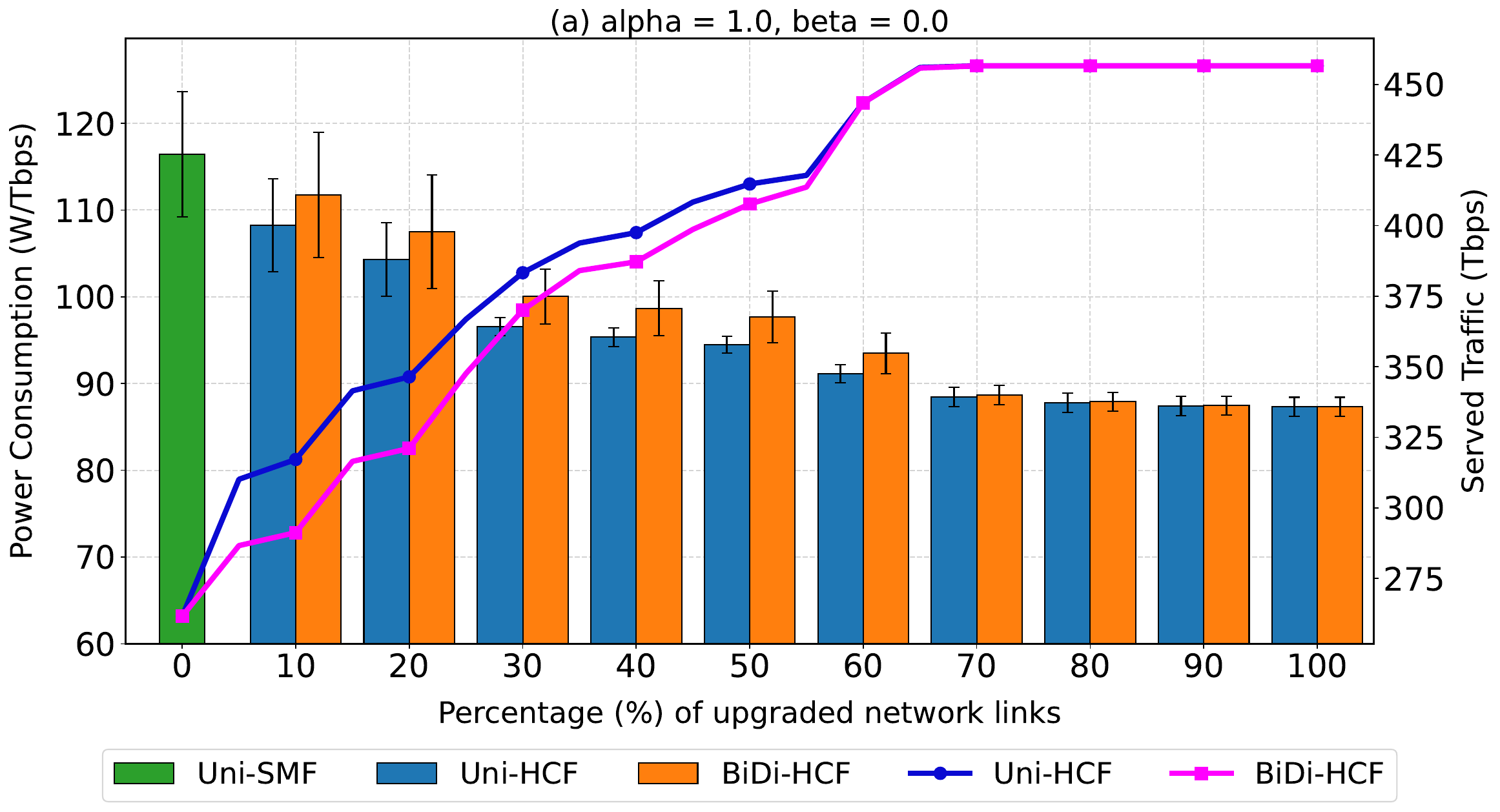}
    \includegraphics[width=0.48\textwidth]{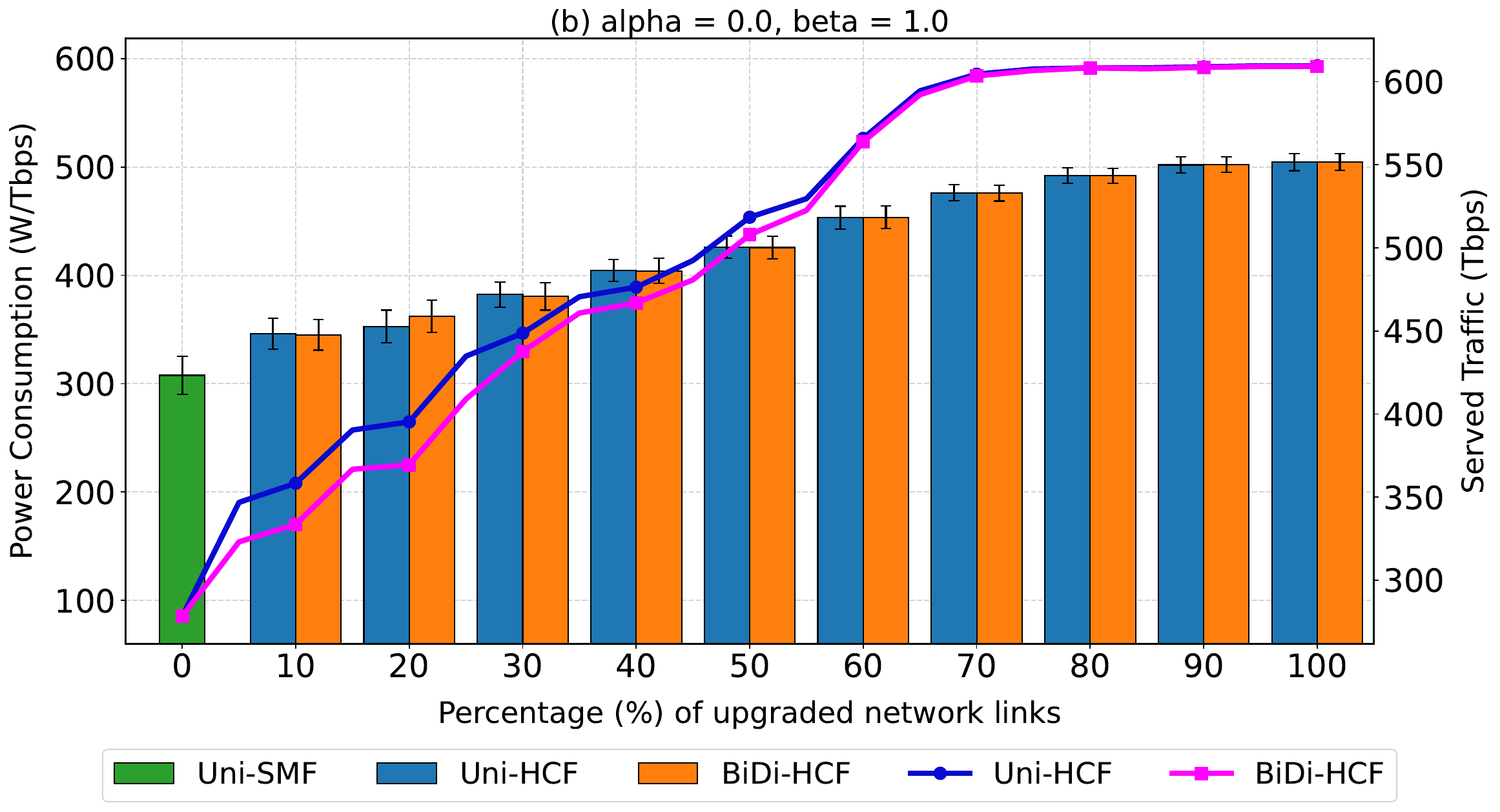}
    \caption{Power consumption per Tbps and total served traffic in Tbps. (a): $\alpha = 1.0, \beta = 0.0$, and (b): $\alpha = 0.0, \beta = 1.0$}
    \vspace{-5pt}
    \label{fig:results}
\end{figure*}

%\newpage
\section{Illustrative Numerical Results}
We evaluate the performance of all five network scenarios  
in a continental-scale EU19 topology (Fig.~8 in \cite{gio_CLS}). 
We generate traffic requests from the set $\{400, 800, 1200, 1600\}$ Gbps with the corresponding probabilities $\{0.25, 0.40, 0.25, 0.10\}$, 
considering incremental traffic until the network blocking probability exceeds 1\%.
We evaluate transponder selection (Eq.~\ref{eq:txp_obj}), considering the transponder set as in Table~\ref{tab:txp-table}, under two weight configurations: $(\alpha=1.0, \beta=0.0)$ and $(\alpha = 0.0, \beta=1.0)$.  
We run 50 independent simulations for each configuration scenario, and report results in terms of power consumption per Tbps (W/Tbps, lower is better), total served traffic (Tbps, higher is better), and percentage of transponder allocation. 

Fig.~\ref{fig:results}(a) shows the transponder power consumption per Tbps for $(\alpha = 1, \beta = 0)$. %, which favors the deployment of energy-efficient ZR+ modules. 
We observe that all HCF-integrated scenarios consistently outperform Uniform Uni-SMF, %(averaging around 115~W/Tbps), 
reaching up to 25\% savings in case of 100\% HCF upgrades.  
Notably, the selective network upgrades (both hybrid Uni-HCF and hybrid BiDi-HCF) lead to a progressive reduction in power consumption as more critical links are upgraded. Focusing on the hybrid BiDi-HCF approach, upgrading (10-50)\% of the network links yields a reduction between (5-20)\% in power consumption compared to Uni-SMF. This is primarily due to the deployment of ZR+ pluggables. 
Moreover, the hybrid BiDi-HCF and Uni-HCF architectures performs similarly well across all upgrade percentages, with performance deviations of less than 3~\%. Qualitatively, this implies that hybrid BiDi-HCF can achieve the same power efficiency while deploying half the number of HCF fiber strands, effectively neutralizing the CapEx and duct-space penalties associated with HCF. Finally, we show that a 100\% network-wide overhaul is unnecessary to minimize power consumption, i.e., upgrading (40-60)\% of network links with BiDi-HCF reduces power consumption to approximately 95~W/Tbps, capturing roughly 85\% of the savings offered by the uniform 100\% HCF scenarios. % (which settle near x~W/Tbps).

Observing the total served traffic in Fig.~\ref{fig:results}(a) (right y-axis), we note an analogous trend. 
The Uni-SMF baseline (Uni-HCF and BiDi-HCF at 0\% HCF links) saturates at less than 280 Tbps, meeting the stopping condition (1\% blocking) much earlier due to limitations in using spectrally-efficient modulation formats. 
Conversely, upgrading to a uniform HCF topology boosts the served traffic to nearly 500 Tbps (a 78\% increase). Similar to power consumption (W/Tbps), the selective BiDi-HCF upgrade demonstrates significant efficiency: upgrading 50\% of network links achieves over 400 Tbps (over 40\% increase) and 
captures 80\% of the gains as in a 100\% network-wide overhaul.

Figure~\ref{fig:results}(b) illustrates the power consumption and total served traffic for $(\alpha = 0, \beta = 1)$. As the network is selectively upgraded with HCF, power consumption per Tbps predictably rises. 
This trend is driven by the prioritization of spectral efficiency, which inherently accommodates higher traffic volumes. % and deploys LH transponders rather than ZR+ pluggables. 
Consequently, overall served traffic increases %surges significantly 
with HCF integration, and BiDi-HCF consistently matches the served traffic of Uni-HCF.
While Uni-SMF exhibits the lowest power consumption, it is constrained by the lowest total served traffic (less than 280 Tbps).
Moreover, we observe that upgrading 50\% of the links with BiDi-HCF increases total served traffic by around 80\% compared to Uni-SMF (while capturing over 80\% of the gains in uniform 100\% HCF overhauls).

\begin{figure}[t]
    \centering    
    \includegraphics[width=1.0\columnwidth]{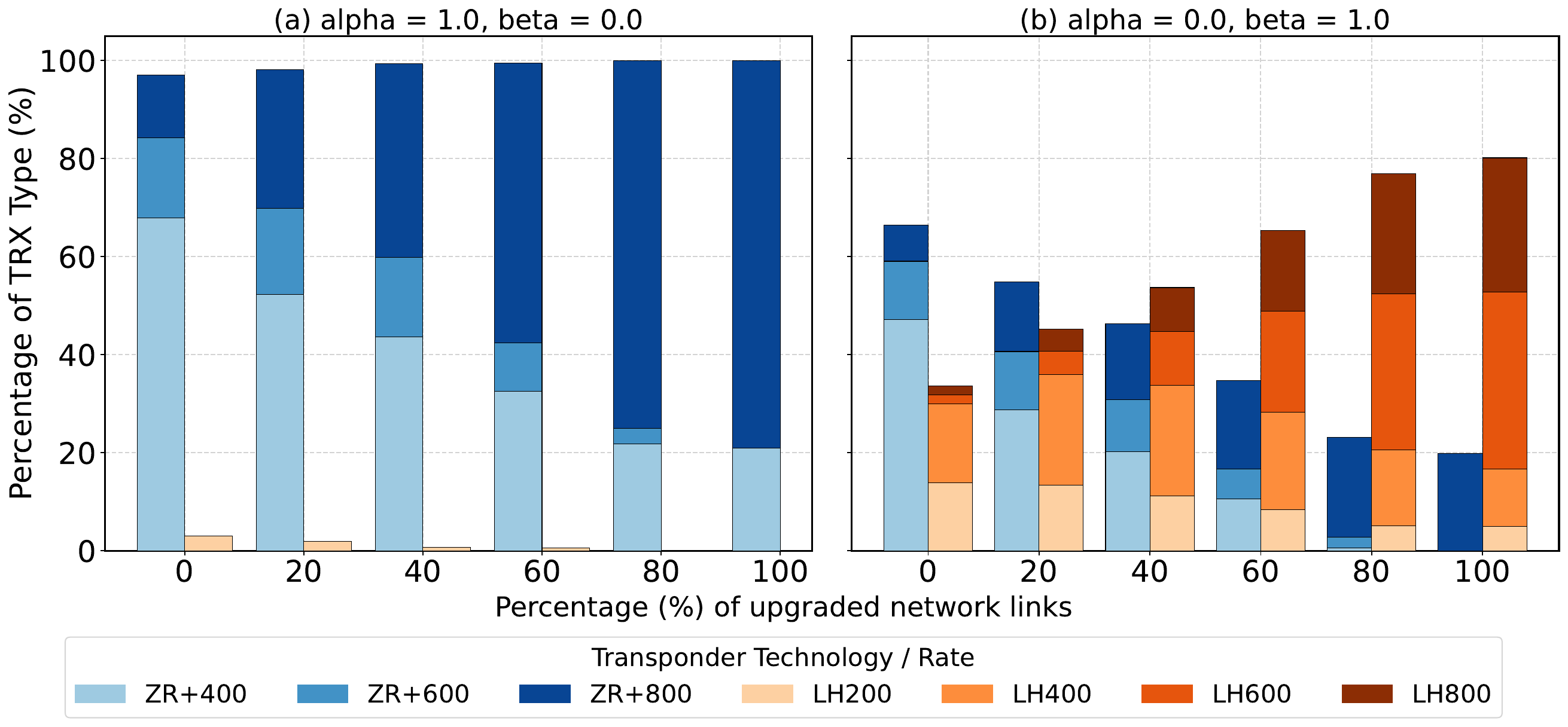}
    \caption{TXP distribution for ($\alpha=1.0, \beta=0.0$) and ($\alpha=0.0, \beta=1.0$) in case of BiDi-HCF}
    \vspace{-10pt}
    \label{fig:txp-distribution}
\end{figure}

The underlying mechanics driving the differences between (a)~$(\alpha = 1, \beta = 0)$ and (b)~$(\alpha = 0, \beta = 1)$ are explained by examining the transponder allocation distributions. 
Fig.~\ref{fig:txp-distribution} shows that when prioritizing power $(\alpha = 1)$, the network favors utilizing energy-efficient ZR+, scaling from 87\% utilization for Uni-SMF to an almost exclusive 99\% utilization for 60\% of network link upgrades, and 100\% in fully upgraded HCF scenarios. Conversely, prioritizing spectral efficiency $(\beta = 1)$ forces the network to rely on LH transponders, which constitute roughly 35\% utilization for Uni-SMF and up to 80\% in a fully upgraded HCF network.

In conclusion, selectively deploying BiDi-HCF in 50\% of network links can increase total served traffic by over 40\% and simultaneously reduce total power consumption per Tbps by over 20\% compared to Uni-SMF. Moreover, a 50\% selective BiDi-HCF upgrade captures over 80\% of the gains inherent to a full 100\% Uni-HCF overhaul.

%-------------------------------------------------- Acknowledgements Section -------------------------------------------------------%
\clearpage
%\section{Acknowledgements}

%-------------------------------------------------- Bibliography Section -------------------------------------------------------%
% see also https://tex.stackexchange.com/questions/55030/text-before-references-but-after-bibliography-title-with-bibtex as of 2024-02-29
\defbibnote{myprenote}{%
%Citations must be easy and quick to find. More precisely:
%\begin{itemize}
%    \item Please list all the authors. 
%    \item The title must be given in full length. 
%    \item Journal and conference names should not be abbreviated but rather given in full length.
%    \item The DOI number should be added incl. a link.
%\end{itemize}
}
\printbibliography[prenote=myprenote]

\vspace{-4mm}

%%%%%%%%%%%%%%%%%%%%%%%%%%%%%%%%%%%%%%%%%%%%%
%---------------------------------------------- End of Document -----------------------------------------------%
\end{document}